\documentclass[aps, prb,twocolumn]{revtex4}

\usepackage{amsmath,amssymb}
\usepackage[utf8]{inputenc}
\usepackage{hyperref}
\usepackage{graphicx}
\usepackage{subfig}

\newcommand{\deriv}[3][1]{\frac{\mathrm{d} \ifnum #1>1 ^{#1} \fi #2}{\mathrm{d} #3 \ifnum #1>1 ^{#1} \fi}}

\newcommand{\secref}[1]{\ref{#1}}
\newcommand{\figref}[1]{\ref{#1}}
\renewcommand{\eqref}[1]{(\ref{#1})}

\begin{document}
\title{Analytical results for the Morse potential in $s$-wave ultracold scattering: three dimensional \emph{vs.} one dimensional problem}

\author{Asaf Paris-Mandoki}
\email{asaf@ciencias.unam.mx}
\author{Roc\'{\i}o J\'auregui}
\email{rocio@fisica.unam.mx}
\affiliation{Instituto de F\'{\i}sica, Universidad Nacional Aut\'onoma de M\'exico}

\begin{abstract}
Taking advantage of the known analytic expression of the eigenfunctions and eigenenergies of the Morse Hamiltonian, explicit expressions are found for the scattering length $a$ and the effective range $r_e$ which determine the $s$-wave scattering of  ultracold atoms. The effects on $a$ and $r_e$ of considering the radial coordinate in the interval $[0,\infty)$ or in the extended region $(-\infty, \infty)$ are studied in detail.
\end{abstract}

\maketitle

\section{Introduction}
Consider a gas of particles so dilute that binary collisions are dominant.
The collision between two particles with thermal momenta $p_{th}=\hbar k_{th} =2\pi/\lambda_{dB}$
is said to be in the ultracold regime if the range of the accompanying interaction 
 is smaller than the de Broglie wave length $\lambda_{dB}$. Under such conditions, as they collide, the particles approach each other more closely than their wavelength, and details of the interaction become blurred. That is, ultracold collisions are expected to be determined by few parameters.

Ultracold collisions may take place in the degenerate limit of a gas of particles for which any particle is permanently within a wavelength from other particles, \emph{i. e.} $\lambda_{dB} > n^{1/3}$ with $n$ the density of the gas.
The search\cite{old}, generation \cite{KCC, KCC2, KCC3, KCC4} and further study \cite{FST1,FST2} of atomic quantum degenerate gases have lead naturally to the analysis of collisions between ultracold atoms.

The purpose of this manuscript is to find analytic expressions  for the $s$-wave parameters that describe such collisions for a potential that exhibits some of the main features of an atom-atom interaction: the Morse potential\cite{morse}
\begin{equation}
V(r) = D((1 - e^{-\beta(r-r_0)})^2 -1), \label{eq:morse-pot}
\end{equation}
where $D$, $\beta$, $r_0$ are positive. This potential is repulsive for short distances, exhibits a local minimum with depth $D$, width determined by $\beta$, located at $r_0$ and is slightly attractive at long distances. Since its proposal it has been extensively used to describe anharmonic features of the vibrational spectra of diatomic molecules.

Simple $s$-wave analytical solutions for the potential can be found for bound\cite{morse,sage} and unbound\cite{matsumoto} states. The key is to use an auxiliary mathematical problem where the radial coordinate $r$,  that is physically constrained to the interval $[0,\infty)$,
is allowed to vary in $(-\infty, \infty)$. In this work we analyze the consequences of using this auxiliary problem instead of the original one derived directly from the Schr\"odinger equation.

In general, for a spherically symmetric potential $V(r)$ and elastic collisions, the scattering effects at any relative momenta $p = \hbar k$ are contained in the partial wave phase shifts $\delta_\ell(k)$. 
It can be shown that as $k \rightarrow 0$, the $s$-wave phase shift $\delta_{\ell=0}(k)$ can be expanded as\cite{blatt,joachain}
\begin{equation}
k\cot\delta_0(k) = -\frac{1}{a}+\tfrac{1}{2}r_ek^2+\cdots;
\label{eq:low_k_exp}
\end{equation}
 $a$ is known as the scattering length and $r_e$ as the effective range. For other partial waves
 $\delta_\ell(k)/k$ goes to zero as $k\rightarrow 0$; $s$-wave collisions contribute to the scattering between bosons and distinguishable particles.  As a consequence, in those cases, collisions in the ultracold regime are expected to be isotropic and characterized by the scattering length.

In this article, expressions for $a$ and $r_e$ are obtained for the Morse Hamiltonian.
We begin in Section \secref{sec:radial} by making a brief revision of the  bound and unbound eigenfunctions of the Morse Hamiltonian that vanish as
$r\rightarrow -\infty$. From those unbound functions, the phase shift $\delta_0(k)$ is explicitly calculated and the scattering parameters $a$ and $r_e$ are written in an analytical closed form. In Section \secref{sec:cost}, we study
the  bound and unbound eigenfunctions of the Morse Hamiltonian
with the boundary condition that nullify $u$ as $r\rightarrow 0$, which is compatible with a radial coordinate restricted to the interval $ [0,\infty)$.
In an analogous way as for the auxiliary problem, the phase shift $\delta_0(k)$ can be calculated and the scattering parameters are implicitly found.
A comparison between the auxiliary and the physical system results is then performed.

\section{Radial solutions for the auxiliary problem\label{sec:radial}}
The Schr\"odinger equation
\begin{equation}
\left[\frac{\hbar^2}{2\mu} \frac{d^2}{dr^2}+E-V(r)\right]u(r) = 0.
\label{eq:radial}
\end{equation}
can be related to the stationary dynamics of a one dimensional collision of two
particles with reduced mass $\mu$ and  Hamitonian eigenenergy $E$, or to  a three dimensional $s$-wave problem for which the radial wavefunction has been written in the form $R(r) =u(r)/r$. Taking $V(r)$ as the Morse potential, Eq.~\ref{eq:morse-pot}, and
introducing the variables $d=\sqrt{2\mu D}/\hbar\beta$, $b = \sqrt{2\mu E}/\hbar\beta$ and $z = 2de^{-\beta(r-r_0)}$ a direct calculation shows that the general solution to Eq.~\eqref{eq:radial} is
\begin{eqnarray}
u_b(z) &=& e^{-z/2}C_1z^{+ib}M(\tfrac{1}{2}+ib-d,1+2ib,z) \nonumber \\
&+&e^{-z/2}C_2z^{-ib}M(\tfrac{1}{2}-ib-d,1-2ib,z),
\end{eqnarray}
where  $C_1$ and $C_2$ are constants to be determined and 
\begin{equation}
M(p,q,z)=\sum_{n=0}^\infty\frac{(p)_nz^n}{(q)_n n!},
\end{equation}
is Kummer's function\cite{Abramowitz1964} with $(p)_n$ the Pochhammer symbol. It will be useful to know that
\begin{equation}
M(p,q,z) = \frac{\Gamma(q) }{\Gamma(p) }e^{z}z^{p-q}\left[1+O\left(|z|^{-1}\right)\right]
\label{eq:M_z_lejos}
\end{equation}
when the real part of $z$ is positive.

\subsection{Bound States}
The bound states are determined by having $E<0$. In this case $b = i\sqrt{2\mu |E|}/\hbar\beta=i|b|$,
and
\begin{eqnarray}
u_b(z) &=& e^{-z/2}C_1z^{-|b|}M(\tfrac{1}{2}-|b|-d,1-2|b|,z) \nonumber \\
&+&e^{-z/2}C_2z^{+|b|}M(\tfrac{1}{2}+|b|-d,1+2|b|,z).
\end{eqnarray}
Since $u_b$ should not diverge when $z\ll 1$ ($\beta r\gg1$), $C_1$ must be zero.
We now need to apply a second boundary condition which will determine the quantization. Solving the 3D radial equation would require demanding $u_b\left (z(r)\right ) = 0$ when $r=0$; however, by applying the condition where $r\rightarrow -\infty$, the wave functions and eigenvalues take a much simpler form which is analytically tractable\cite{morse}. We will analyze the consequences of using this method in section \ref{sec:cost}.
When $r\rightarrow -\infty$, $z\rightarrow\infty$ and by using Equation~\eqref{eq:M_z_lejos} it is found that
\begin{equation}
u_b(z)=C_2 e^{z/2}z^{-\frac{1}{2}-d}\frac{\Gamma\left(1+2|b|\right)}{\Gamma\left(\frac{1}{2}+|b|-d\right)} \left[1+O\left(|z|^{-1}\right)\right].
\end{equation}
It is worth noting that $u_b(z)$ grows exponentially as $z$ grows unless $1/2+|b|-d$ is a negative integer. Since  $u_b(z)$ should not grow exponentially in that region we define $-n=1/2+|b|-d$, where $n$ is a positive integer. This condition determines the quantization of the energy levels, $b_n=|b|=d-n-1/2$, $n\in{0,1,2,\dots}$ or
$E_n = -D+\hbar\beta\sqrt{2D/\mu}\left(n+1/2\right)-(\hbar^2\beta^2/2\mu)\left(n+1/2\right)^2$. Since $b_n$ is always positive then $n$ can only take a finite number of values for a given $d$. This means that the Morse potential can only hold a finite number of bound states. Since the first argument of $M$ turns out to be an integer,  the solution can be rewritten in terms of Laguerre polynomials as
\begin{equation}
u_n(z) = \left( \frac{\beta n!\, 2|b_n|}{\Gamma\left(2|b_n|+n+1\right)}\right)^{1/2} e^{-z/2} z^{|b_n|} L_n^{(2|b_n|)}(z),
\label{eq:sol_ligada}
\end{equation}
with which the bound solutions for $l=0$ are fully determined.

\subsection{Unbound States}
The unbound states are determined by having $E>0$. In this case $b =  \sqrt{2\mu |E|}/\hbar \beta=|b|$,
and
\begin{eqnarray}
u_b(z) &=& e^{-z/2}C_1z^{+i|b|}M(\tfrac{1}{2}+i|b|-d,1+2i|b|,z) \nonumber \\
&+&e^{-z/2}C_2z^{-i|b|}M(\tfrac{1}{2}-i|b|-d,1-2i|b|,z).
\end{eqnarray}
Again, we apply a boundary condition when $r\rightarrow -\infty$ which means that $z\rightarrow \infty$ where we require that \mbox{$u_b(z)\rightarrow 0$}. Using Equation~\eqref{eq:M_z_lejos} it is found that
\begin{eqnarray}
u_b(z) &=& e^{z/2}z^{-\frac{1}{2}-d} \nonumber \\
&\times & \left[C_1 \frac{\Gamma(1+2i|b|)}{\Gamma(\tfrac{1}{2}+i|b|-d)}+C_2 \frac{\Gamma(1-2i|b|)}{\Gamma(\tfrac{1}{2}-i|b|-d)}\right] \nonumber\\
&\times & \left(1+O(|z|^{-1})\right).
\end{eqnarray}
As in the bound case $u_b(z)$ grows exponentially with $z$ and we can only play with $C_1$ and $C_2$ to satisfy the condition $u_b(z(r))\rightarrow 0$ as $r\rightarrow -\infty$. For this we look for the relationship between $C_1$ and $C_2$ that nullifies the factor with square parenthesis and find that
\begin{equation}
\frac{C_1}{C_2} = \frac{\Gamma(-2i|b|)}{\Gamma(\tfrac{1}{2}-i|b|-d)}\overline{\left(\frac{\Gamma(\tfrac{1}{2}-i|b|-d)}{\Gamma(-2i|b|)}\right)},
\end{equation}
where $\overline{s}$ means the complex conjugate of $s$. Therefore we define $A(b) = \Gamma(-2i|b|)/\Gamma(\tfrac{1}{2}-i|b|-d)$, so we can satisfy the condition with $C_1 = \tilde{C}_b A(b)$ and $C_2 = \tilde{C}_b \overline{A}(b)$, where $\tilde{C}_b$ is a normalization factor that can depend on $b$. In this manner, the solution has the form\cite{matsumoto}
\begin{eqnarray}
u_b(z) &=& 2 e^{-z/2} \tilde{C}_b \nonumber \\
&\times & \Re\left\{A(b)z^{i|b|}M (\tfrac{1}{2}+i|b|-d,|+2i|b|,z )\right\},
\label{eq:sol_libre}
\end{eqnarray}
where $\Re$ means the real part.

It is important to analyze the asymptotic behavior of the solutions since the scattering phase shift depends on this. When $r \rightarrow \infty$, $ z \rightarrow 0 $ in such way that Equation~\eqref{eq:sol_libre} simplifies to
\begin{equation}
u_b(z) \mathop{\rightarrow}_{z\rightarrow 0} 2 \tilde{C}_b \Re\left\{A(b)z^{i|b|}\right\}.
\end{equation}
Writing it in terms of $r$ we get that the asymptotic behavior is given by
\begin{eqnarray}
u_b(z) &\approx& 2 \tilde{C}_b\Re\left\{A(b)\left(2de^{\beta r_0}\right)^{i|b|}\right\}\cos(kr) \nonumber \\
&+&2 \tilde{C}_b\Im\left\{A(b)\left(2de^{\beta r_0}\right)^{i|b|}\right\}\sin(kr),
\label{eq:asymptotic}
\end{eqnarray}
where $k=|b|\beta$ is the asymptotic wave number and $\Im$ means the imaginary part.

In absence of a potential, the normalized radial wave function has the form $u_k(r)= \sqrt{2/\pi} \sin(kr)$.
The presence of the Morse potential also produces an asymptotic sinusoidal solution as seen in Equation~\eqref{eq:asymptotic}. However, the cosine term results in an $s$-wave phase shift for the auxiliary problem $\delta_0^{(aux)}(k)$ which satisfies
\begin{equation}
\tan\delta_0^{(aux)}(k) = \frac{\Re\left\{A(k/\beta)\left(2de^{\beta r_0}\right)^{ik/\beta}\right\}}{\Im\left\{A(k/\beta)\left(2de^{\beta r_0}\right)^{ik/\beta}\right\}}.
\end{equation}
On the other hand $\Re\{s\}/\Im\{s\} = \tan(\arg(i\overline{s}))$, so
\begin{eqnarray}
\delta_0^{(aux)}(k) &=& \arg\left(i\overline{\left(A(k/\beta)\left(2de^{\beta r_0}\right)^{ik/\beta}\right)}\right) \nonumber \\
&=& \frac{\pi}{2}-\arg A(k/\beta)-\frac{k}{\beta}\log(2d)-kr_0
\end{eqnarray}
modulo $\pi$. Moreover,
\begin{equation}
\arg A\left (\tfrac{k}{\beta}\right ) = \arg\Gamma\left(-i\tfrac{k}{\beta}\right)-\arg\Gamma\left(\tfrac{1}{2}-i\tfrac{k}{\beta}-d\right).
\end{equation}
Using an expansion for $\arg\Gamma(x+iy)$\cite{Abramowitz1964} we finally get the phase shift
\begin{equation}
\delta_0^{(aux)}(k) = -\frac{k}{\beta}\left(\gamma+\ln(2d)+\beta r_0\right)+\Xi,
\label{eq:phaseshift}
\end{equation}
where $\gamma$ is the Euler-Mascheroni constant and
\begin{equation}
\Xi = \sum_{n=1}^\infty\frac{k}{\beta n}-\arctan\frac{2k}{\beta n}+\arctan\frac{k}{\beta\left(n-d-\frac{1}{2}\right)}.
\label{eq:phaseshift_sum}
\end{equation}
At this point we have full knowledge of the $s$-wave scattering phase shift from which, in principle, we can extract all the $s$-wave scattering information for the Morse potential, as we will exemplify when we calculate the scattering length and effective range.

We will now proceed by calculating the normalization factor. Using the scattering phase shift, we write the asymptotic behavior as
\begin{equation}
u_k(r)\mathop{\rightarrow}_{r \rightarrow \infty  } 2 \tilde{C}_b |A(k/\beta)|\sin\left(kr+\delta_0(k) \right).
\end{equation}
Following Bethe,\cite{bethe} the normalization of continuum states is defined by their asymptotic behavior in which we require that
\begin{equation}
u_k(r) \mathop{\rightarrow}_{r \rightarrow \infty  } \sqrt{\frac{2}{\pi	} } \sin(kr+\delta_0(k)).
\end{equation}
Therefore, the normalization factor $\tilde{C}_b$ is given by
\begin{eqnarray}
\tilde{C}_b &=& \frac{1}{\pi}\left(\frac{|b|\sinh 2\pi |b|}{\nu^2+|b|^2}\right)^{1/2}e^{\gamma \nu} \\
&\times& \prod_{n=1}^{\infty}\left[\left(1-\frac{\nu}{n}\right)^2+\left(\frac{|b|}{ n}\right)^2 \right]^{-1/2}e^{-\nu/n},
\end{eqnarray}
where $\nu=d-1/2$.

To finalize this section we analyze the low energy scattering behavior of the Morse potential. As the particles energy tends to zero, the $s$-wave scattering amplitude determined by $\delta_0$ becomes dominant.  For low energy, Eq.~\eqref{eq:low_k_exp} defines $a$, the \emph{scattering length} and $r_e$, the \emph{effective range}. In order to find expressions for them we will first calculate the low energy behavior of $\delta_0(k)$ and afterwards write the expansion~\eqref{eq:low_k_exp}. By identifying the coefficients of the expansion we will find the parameters we seek.
We begin by using the fact that $\arctan x = x-\frac{x^3}{3}+O(x^5)$ to rewrite the series~\eqref{eq:phaseshift_sum} when $k/\beta\ll 1$ as

\begin{eqnarray}
\Xi &=& -\left[\psi(\tfrac{1}{2}-d) + \gamma\right]\frac{k}{\beta} \nonumber \\
&+&\left[\tfrac{1}{6}\psi^{(2)}(\tfrac{1}{2}-d)+\tfrac{8}{3} \zeta(3) \right]\left( \frac{k}{\beta}\right)^3+O\left(k^5\right ),
\end{eqnarray}
where $\psi^{(n)}$ is the polygamma function\cite{Abramowitz1964} and $\psi=\psi^{(0)}$. Defining two variables
\begin{equation}
\eta = \frac{1}{\beta}\left ( 2\gamma+\ln(2d) +\beta r_0+\psi(\tfrac{1}{2}-d)\right )
\end{equation}
and
\begin{equation}
\xi = \frac{1}{\beta^3}\left (\tfrac{1}{6}\psi^{(2)}(\tfrac{1}{2}-d)+\tfrac{8}{3}\zeta(3)\right ),
\end{equation}
the phase shift is rewritten as
\begin{equation}
\delta^{(aux)}_0(k) =-k\eta+k^3\xi +O\left(k^5 \right).
\end{equation}
On the other hand, using the Maclaurin expansion of $\cot\delta_0$ we write,
\begin{equation}
k\cot\delta_0(k) = \frac{k}{\delta_0(k)}-\frac{k\delta_0(k)}{3}+O\left(\delta_0^3 \right),
\end{equation}
which yields
\begin{equation}
k\cot\delta_0(k) = -\frac{1}{\eta}+k^2\left( \frac{\eta}{3}-\frac{\xi}{\eta^2}\right) +O\left(k^3\right ).
\end{equation}
Identifying the terms in the previous expression with the ones in Eq.~\eqref{eq:low_k_exp} we find that the scattering length is given by
\begin{equation}
a = r_0+\frac{1}{\beta}\left[ 2\gamma+\ln(2d)+\psi(\tfrac{1}{2}-d)\right],\label{eq:alpha}
\end{equation}
while the effective range is
\begin{equation}
r_e = \frac{2}{3}a-\frac{\psi^{(2)}(\tfrac{1}{2}-d)+16\zeta(3)}{3\beta^ 3a^ 2} \label{eq:re}.
\end{equation}

\begin{figure}
\includegraphics[width=0.7\linewidth]{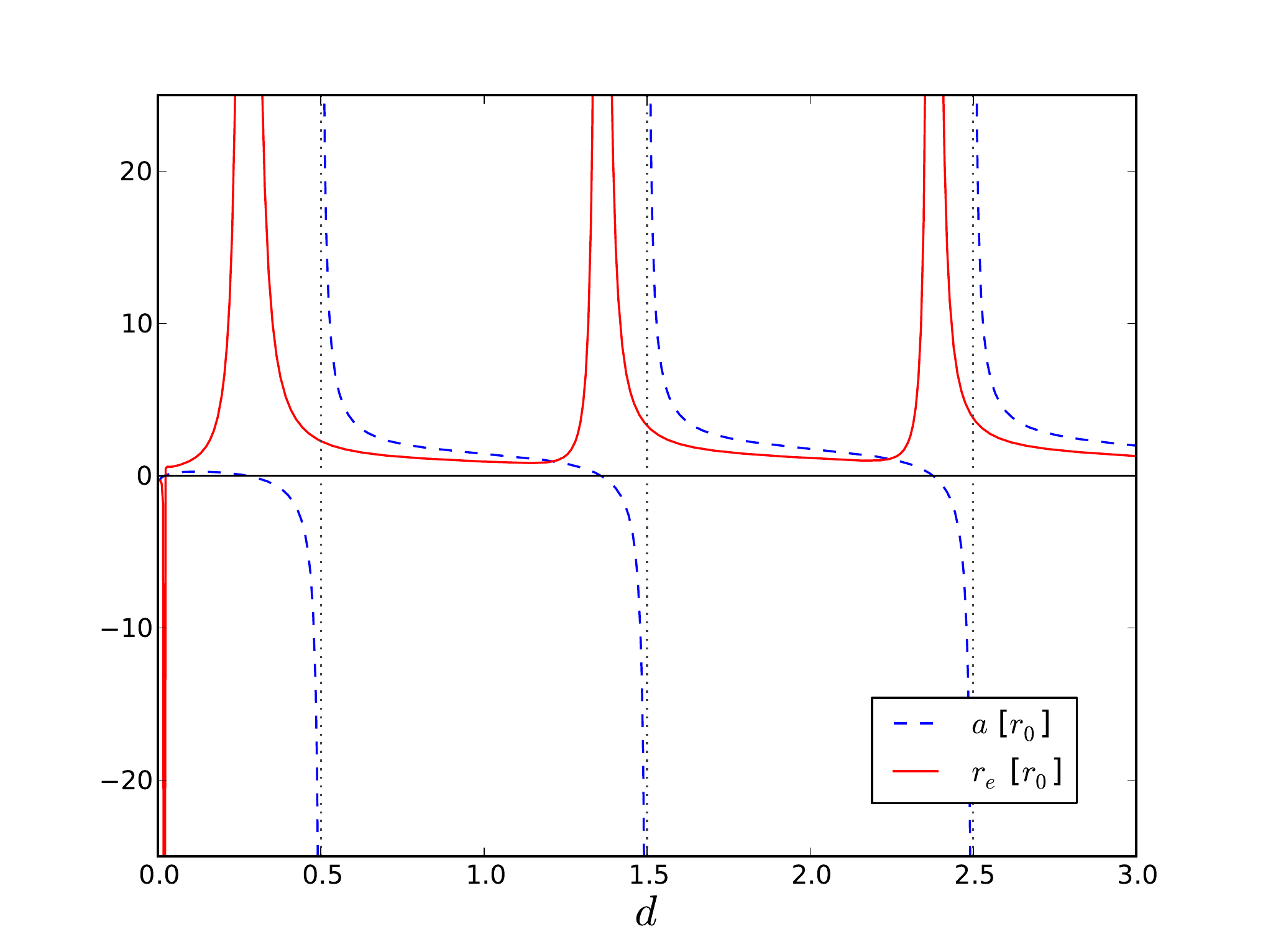}
\caption{(Color online) The scattering length and effective range obtained from the auxiliary problem as a function of $d=\sqrt{2\mu D}/\hbar\beta$ for $r_0\beta=4.15$.}
\label{fig:resultado}
\end{figure}

The scattering length and effective range as a function of the depth of the potential are illustrated in Figure~\figref{fig:resultado} for the case $r_0\beta=4.15$ which could correspond to two atoms of ${}^6Li$ colliding with an electronic state ${}^3\Sigma_u^+$ when $D=40$meV \cite{lithium}.
One notices that for $d\approx n+1/2$, $n=0,1,2,\dots$ the scattering length is not well defined since
its limiting value from the right would be negative and diverging while from left it would be positive and diverging.
 In the nomenclature of scattering theory that condition is known as the unitarity limit or the zero energy resonance\cite{joachain}. For those values of
 $d$ the Morse potential is about to support a new bound state.

 As for the effective range, it is always positive with the exception $d\ll 1$. This condition is not shared by other potentials like the square well which admit positive and negative values of $r_e$ for extended regions of the potential depth. 
We also observe that the $r_e$ resonances are located to the left of the $a$ resonances where $a$ becomes zero.

\section{Radial solutions for the physical problem: consequences of including $r<0$ in the auxiliary problem\label{sec:cost}}
An auxiliary mathematical problem was used to find the analytical results shown in the previous sections. The purpose of this section is to understand  better the trade-offs of replacing the physical problem by the auxiliary one.

\subsection{Bound States}

First of all, the general methodology, described at the beginning of last section, when applied to the auxiliary problem yields simple analytical solutions. Nevertheless, if we allow $r$ to vary only in the $[0,\infty)$ interval and demand $u_b(z(r))|_{r=0} = 0$  that methodology also yields analytical solutions that do not reduce to the simple expression~\eqref{eq:sol_ligada}.
 These solutions are
\begin{equation}
u_j(z) = C e^{-z/2}z^{+|b_j|}M(\tfrac{1}{2}+|b_j|-d,1+2|b_j|,z).
\end{equation}
Here $\vert b_j\vert $, which determine the eigenenergies, are the positive roots of the equation in $b$ given by
\begin{equation}
M\left(\tfrac{1}{2}+|b|-d,1+2|b|,2de^{\beta r_0}\right ) =0.
\end{equation}

\begin{figure}
\includegraphics[width=0.7\linewidth]{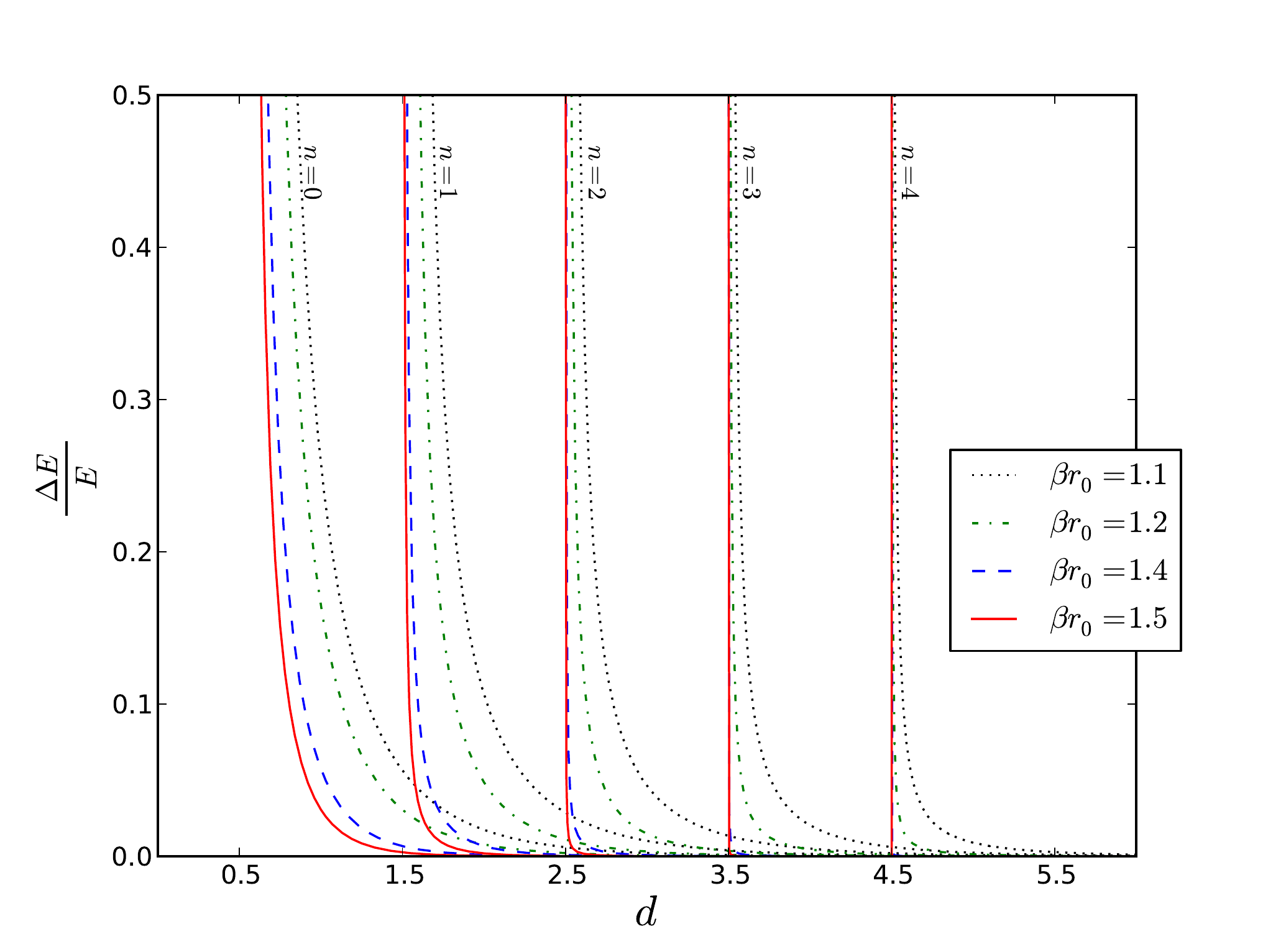}
\caption{(Color online) Comparison between the energy values that result in the physical and auxiliary problems \mbox{($\Delta E = E^{(aux)}-E$)} as a function of
the scaled potential depth $d=\sqrt{2\mu D}/\hbar\beta$ and several values of the product $r_0 \beta$, with $\beta$ the inverse of the potential range and
$r_0$  the equilibrium distance of the potential.}
\label{fig:energias}
\end{figure}

In Figure~\figref{fig:energias} we compare the energy eigenvalues that result in the physical and auxiliary problems
for several values of the product $r_0 \beta$. As noticed before, if $d$ is in the interval $[n+1/2, n+3/2)$ for
$n=0,1,2,...$, then precisely $n+1$ bound states are supported for the auxiliary Morse problem. For the physical problem, this occurs
for greater values of $d$. This effect is more evident for  small values of $d$ and $r_0\beta$. For instance,
for $r_0\beta \sim 1$ the Morse potential supports no bound states until  $d>0.6$. For $r_0\beta \sim 4$, the first bound state is
found for $d>0.5$ which is the same value (modulo the limited double precision of the computer calculation) found in the real problem.

\subsection{Unbound States}

\begin{figure}
\includegraphics[width=0.7\linewidth]{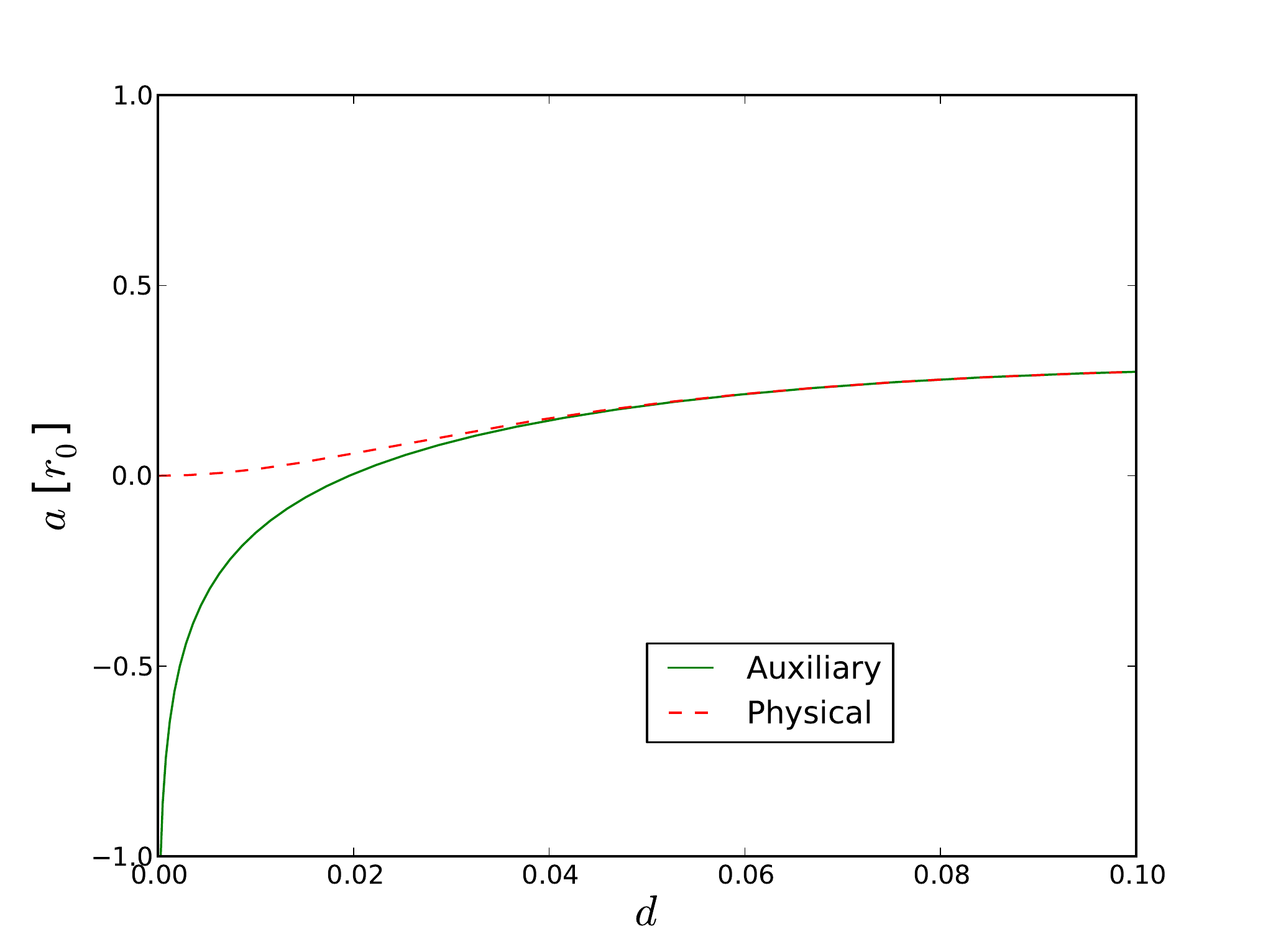}
\caption{(Color online) The scattering length  as a function of $d=\sqrt{2\mu D}/\hbar\beta$ for $r_0\beta=4.15$ for the physical and auxiliary problems.}
\label{fig:efecto_alfa}
\end{figure}

For small potential depths, $d \rightarrow 0$, the scattering length evaluated using Eq.~\eqref{eq:alpha}
exhibits a logarithmic divergence and the free-particle expresions are not obtained. In order to verify the physical reliability  of this property,
 $a$ must be evaluated  considering the physical boundary condition $u_b(z(r=0))=0$.
  Imposing it, a direct calculation shows that the radial functions now take the form
\begin{equation}
u_b(z) = 2i\tilde C_b e^{-z/2} \Im\left\{\tilde{A}(b)z^{i|b|}M\left(\tfrac{1}{2}+i|b|-d,1+2i|b|, z\right) \right\},
\end{equation}
with
\begin{equation}
\tilde{A}(b) = z_0^{-i|b|}M\left(\tfrac{1}{2}-i|b|-d,1-2i|b|, z_0\right),
\end{equation}
and $ z_0 = 2de^ {\beta r_0}$.
In a similar way as we obtained the phase shift before we now get
\begin{eqnarray}
\delta_0(b) &=& -\arg\left[z_0^{i|b|}\tilde{A}(b)\right]\nonumber\\
& =& -\arg M\left(\tfrac{1}{2}-i|b|-d,1-2i|b|, z_0\right).
\label{eq:fas_cortar_en_cero}
\end{eqnarray}
Notice that, due to the structure of $\tilde A (b)$, the factor $z_0^{i|b|}$ that appears in the phase shift $\delta_0$ (which
gives rise to the divergence of $a$ in Eq.~\eqref{eq:alpha} is now canceled. From Eq.~\eqref{eq:fas_cortar_en_cero}
$a$ can be calculated by performing numerically the limit $k\rightarrow 0$ of $k\cot\delta_0(k)$, Eq.~\eqref{eq:low_k_exp}. In Fig.~\figref{fig:efecto_alfa} the resulting scattering lengths are illustrated and one can see that the divergence when $d\rightarrow 0$ is removed for the real problem. As for the effective range,
it now becomes zero as $d\rightarrow 0$ and for larger values of $d$ the differences between the values of $r_e$ for the physical and auxiliary problem result to be less than one percent for all the studied cases.

\section{Conclusions}

  In this work, analytic expressions have been obtained that solve the eigenvalue problem of the Morse Hamiltonian under two different
  boundary conditions. This Hamitonian is widely used to model the $s$-wave anharmonic vibrations of  nuclei in diatomic molecules
  and supports a finite number of bound states.
  It was shown that the eigenvalue of the highest excited bound state derived from the boundary condition $u(z(r\rightarrow-\infty))=0$
   differs significantly from that derived from the condition $u(z(r=0))=0$ for potentials with a range similar to the equilibrium position,
   $\beta r_0\approx 1$, at the unitarity limit, $i.$ $e.$ with a  potential depth $D$ close to the values that
   yield the possibility for the Hamiltonian to support a new bound state. Outside this limit, the difference between the energy eigenvalues for the auxiliary
   and the physical boundary condition becomes small. This is congruent with using the former in the standard analysis of molecular vibrations.

   We also derived analytical expressions for the phase shift in binary collisions both for the auxiliary and the physical problem.
   From them, the most important parameters necessary to describe an ultracold collision, that is, the scattering length $a$ and
   effective range $r_e$ were evaluated. A divergence of $a$ predicted for very small potential depths $d=\sqrt{2\mu D}/\hbar\beta \ll 1$
   for the auxiliary problem was removed by imposing the physical boundary condition.
  This analysis illustrates the fact that, even though the scattering length is a property that summarizes the asymptotic
   behavior of a wave function at $r\rightarrow \infty$,
  it is highly influenced by its behavior  at the origin. It is important to mention that precisely this observation is
  the basis of the theories that use effective potentials to incorporate scattering effects. Perhaps the best well known example of the latter
  is the Gross-Pitaevskii equation \cite{gross,pitaevskii} that models an ultracold gas of bosons.

{\bf Acknowledgement.} We acknowledge partial support by DGAPA-UNAM through the project IN111109.

\end{document}